\documentclass[a4paper,10pt]{article}
\usepackage[utf8]{inputenc}

\usepackage{amsfonts}
\usepackage{amsmath}
\usepackage{amsthm}
\usepackage{graphicx}
\usepackage{pstricks}
\usepackage{color}
\usepackage{stmaryrd}
\usepackage{amsfonts}
\usepackage{amsmath}
\usepackage{stmaryrd}
\usepackage{multicol}
\usepackage{subfig}
\usepackage{pst-node}

\newcommand{\sgn}{\mathop{\mathrm{sgn}}}
\newcommand{\ljump}{\llbracket}
\newcommand{\rjump}{\rrbracket}

\newcommand\bx{\mbox{\boldmath $x$}}

\newcommand\bn{\mbox{\boldmath $n$}}
\newcommand\bs{\mbox{\boldmath $\sigma$}}
\newcommand\bu{\mbox{\boldmath $u$}}
\newcommand\bp{\mbox{\boldmath $p$}}
\newcommand\bS{\mbox{\boldmath $\Sigma$}}
\newcommand\bU{\mbox{\boldmath $U$}}
\newcommand\bR{\mbox{\boldmath $R$}}
\newcommand\bI{\mbox{\boldmath $I$}}
\newcommand\bE{\mbox{\boldmath $E$}}

\newcommand\bC{\mbox{\boldmath $C$}}
\newcommand\bD{\mbox{\boldmath $D$}}

\newcommand{\bmA}{\mbox{\boldmath $\mathcal{A}$}}
\newcommand{\bmB}{\mbox{\boldmath $\mathcal{B}$}}

\newcommand{\jump}[1]{\ljump #1 \rjump}
\newcommand{\av}[1]{\langle #1 \rangle}

\newcommand{\mF}[3]{\mathcal{F}_{#1}^{#2}\left\{#3\right\}}

\newcommand{\mS}{\mathcal{S}}

\title{Interfacial crack integral identities incorporating mean displacement}
\author{Adam Vellender}
\date{}
\begin{document}

\maketitle
\begin{abstract}
 A semi-infinite crack loaded by a general asymmetric system of forces in an infinite bi-material plane is considered. A boundary integral formulation is derived using the fundamental reciprocal identity (Betti formula). The resulting singular integral equations link the applied loading and the full resulting crack displacement profile (not just the displacement jump across the crack). When used in conjunction with previously derived identities, the new identities allow for the full displacement profile to be derived from crack face loadings and vice versa.
\end{abstract}

%
\section{Introduction}
In the modelling of fracture mechanics of dissimilar bodies, singular integral equations play a very important role. Since the early work of Muskhelishvili \cite{muskhelishvili}, singular integral formulations have typically been derived via a Green's function approach. Such an approach introduces difficulties: the resulting integrals are often numerically challenging to compute and the approach requires the loadings applied on crack faces to be symmetric.

Recently, much attention has been paid to the derivation in sufficiently geometrically regular domains of integral equations relating applied loadings on crack faces and the resulting crack opening displacements. This approach hinges on the use of integral transforms (typically Fourier transforms), Betti's reciprocal theorem and weight functions (singular non-trivial solutions of the homogenous traction free problem). It allows for asymmetric balanced loadings and has been used to derive identities corresponding to cracks in dissimilar isotropic \cite{picc2d3d}, anisotropic \cite{morini2013}, and thermodiffusive \cite{morini2015} bimaterials, including those joined by non-ideal interfaces \cite{mishurisv2014}. Given a set of balanced loadings, the integral identities allow for the displacement jump across the crack to be quantified. Such identities can find applications in multiphysics where the elastic problem is coupled with other physical phenomena such as in hydraulic fracture settings where elasticity is coupled with fluid dynamics.

The displacement jump across a fracture described by existing identities do not alone give the full displacement profile however since the resulting displacement is not in general symmetrical. Further identities are therefore required to fully describe the displacement of the crack faces which include the mean as well as the jump in displacement. The aim of the existing paper is to derive such identities and to illustrate their use to solve a wider range of problems.

\subsection{Problem formulation and notation}
We consider an infinite bi-material plane whose two materials are joined via an ideal interface positioned along the $x_1$-axis. A semi-infinite crack is placed at the interface occupying the line
$\Gamma = \{(x_1,x_2): x_1 < 0, x_2 = 0\}$. We refer to the half-planes above and below the interface as $\Pi^{(1)}$ and $\Pi^{(2)}$, respectively. The material
occupying $\Pi^{(j)}$ has shear modulus $\mu_j$ and Poisson's ratio $\nu_j$ for $j = 1,2$.

Ahead of the crack ($x_1>0$), the ideal interface satisfies continuity of both displacement and stresses:
\begin{align}
 \bu(x_1,0^+)-\bu(x_1,0^-)&=0,\\
 \bs_2(x_1,0^+)-\bs_2(x_1,0^-)&=0,
\end{align}
where $\bs_2=(\sigma_{21},\sigma_{22},\sigma_{23})^T$ denotes the traction vector and $\bu=(u_1,u_2,u_3)^T$ the displacement field.
The crack faces ($x_1<0$) are loaded by a system of distributed forces 
\[
 \bs_2(x_1,0^+)=\bp_+^{(-)}(x_1), \qquad \bs_2(x_1,0^-)=\bp_-^{(-)}(x_1).
\]

We will extensively consider the symmetric and skew-symmetric parts of the loadings, which we respectively denote
\begin{equation}
 \av{\bp}^{(-)}=\frac{1}{2}(\bp_+^{(-)} + \bp_-^{(-)}),\qquad \jump{\bp}^{(-)}=\bp_+^{(-)}-\bp_-^{(-)}.
\end{equation}
Here $^{(-)}$ indicates a function is only non-zero on the negative half-line and we have used common notations to denote the average, $\av{f}$, and the jump, $\jump{f}$, of a function $f$ across the plane containing the crack, $x_2=0$. Precisely:
\begin{align}
 \av{f}(x_1)&=\frac{1}{2}\left(f(x_1,0^+)+f(x_1,0^-)\right),\\
 \jump{f}(x_1)&=f(x_1,0^+)-f(x_1,0^-).
\end{align}

\section{Weight functions and the additive Betti formula.}\label{section:betti}
In this section, we use a Betti formula and weight function approach to derive integral identities which relate loadings $\av{\bp}^{(-)}$ and $\jump{\bp}^{(-)}$, displacement jump $\jump{\bu}^{(-)}$, interfacial traction $\av{\bs}^{(+)}$, and mean displacement $\av{\bu}$. Previous approaches have neglected to consider $\av{\bu}$ which is in general nonzero along both positive and negative half-lines. The resulting identities will allow not only for $\av{\bu}$ to be determined for prescribed loadings, but also to recover loadings $\av{p}$ and $\jump{p}$ for given crack displacement profiles.

We will make use of weight functions derived by Piccolroaz \cite{picc2009} which correspond to the presently considered geometry. These, as defined generally by Bueckner \cite{bueckner}, are non-trivial singular solutions of the homogenous traction-free problem. As per Willis and Movchan \cite{willismovchan}, the weight function is defined in a mirrored domain in which the crack occupies $x_2>0$.

The weight functions $\jump{\bU}$ and  $\av{\bU}$, and corresponding traction $\av{\bS}$ are represented by 2$\times$2 matrices. For an elastic bi-material plane there are two linearly independent weight functions 
$\bU^j=[U_1^j,U_2^j]^T,$ $\bS^j=[\Sigma_{1}^j,\Sigma_{2}^j]^T$ for $j=1,2,$ and the weight function tensors can be constructed by arranging the weight function components in columns of 2$\times$2 matrices \cite{picc2009}:
\begin{equation}
 \bU=\begin{bmatrix}
    U_1^1&U_1^2\\
    U_2^1&U_2^2
   \end{bmatrix},
   \qquad
 \bS=\begin{bmatrix}
      \Sigma_1^1 & \Sigma_1^2\\
      \Sigma_2^1 & \Sigma_2^2
     \end{bmatrix}.
\end{equation}

%

Suppose that two displacement fields $\bu^{(1)}$ and $\bu^{(2)}$ with corresponding stress states $\bs^{(1)}$ and $\bs^{(2)}$ satisfy the equations of equilibrium within a region $\Omega$ which is enclosed by $\partial\Omega$. The Betti formula then takes the form
\begin{equation}
 \int\limits_{\partial \Omega}\left\{\bs^{(1)}\bn\cdot\bu^{(2)}-\bs^{(2)}\bn\cdot\bu^{(1)}\right\}\mathrm{d}s=0,
\end{equation}
where $\bn$ is the outward normal to $\partial\Omega$. Applying the Betti formula to the physical fields and to weight functions for the upper and lower half plane respectively gives the following two expressions.
\begin{equation}\label{chisel:bettitop}
 \int\limits_{-\infty}^\infty{\left\{\bR\bU(x_1'-x_1,0^+)\bs(x_1,0^+)-\bR\bS(x_1'-x_1,0^+)\bu(x_1,0^+)\right\}\mathrm{d}x_1}=0,
\end{equation}
\begin{equation}\label{chisel:bettibottom}
 \int\limits_{-\infty}^\infty{\left\{\bR\bU(x_1'-x_1,0^-)\bs(x_1,0^-)-\bR\bS(x_1'-x_1,0^-)\bu(x_1,0^-)\right\}\mathrm{d}x_1}=0,
\end{equation}
where $R$ is the rotation matrix given below along with matrices $\bI$ and $\bE$ which are to be used later: 
\begin{equation}
 \bR=\begin{bmatrix}
      -1 & 0\\
      0 & 1
     \end{bmatrix}, \qquad
\bI=\begin{bmatrix}
      1 & 0\\
      0 & 1
     \end{bmatrix}, \qquad
\bE=\begin{bmatrix}
  0 & 1\\
    -1 & 0
\end{bmatrix}. \qquad
\end{equation}
These identities were proved under the assumption that the integrand decays faster at infinity than $1/R$ along any ray. Typically these expressions are subtracted from one another; this yields expressions involving $\jump \bu$ but neglects $\av{\bu}$. Instead summing (\ref{chisel:bettitop}) and (\ref{chisel:bettibottom}) yields the expression
\begin{align}
  \int\limits_{-\infty}^\infty\{&\bR\bU(x_1'-x_1,0^+)\bs(x_1,0^+)+\bR\bU(x_1'-x_1,0^-)\bs(x_1,0^-)\nonumber\\
&-[\bR\bS(x_1'-x_1,0^+)\bu(x_1,0^+)+\bR\bS(x_1'-x_1,0^-)\bu(x_1,0^-)]\}\mathrm{d}x_1=0.\label{chisel:bettishort}
\end{align}
We split the terms for physical stress into two parts (along the crack and along the interface), writing
\begin{equation}\label{plane:splitting}
 \bs(x_1,0^\pm)=\bp_\pm^{(-)}(x_1)+\av{\bs}^{(+)}(x_1),
\end{equation}
where $\bp_\pm^{(-)}$ and $\av{\bs}^{(+)}$ are defined as follows
\begin{equation}
 \bp_\pm^{(-)}(x_1)=\mathcal{H}(-x_1)\bs(x_1,0^\pm),\qquad
\av{\bs}^{(+)}(x_1)=\mathcal{H}(x_1)\av{\bs}(x_1,0),
\end{equation}
and $\mathcal{H}(x_1)$ denotes the Heaviside step function. The functions $\bp_\pm^{(-)}(x)$ represent the loadings on the crack faces. Expression (\ref{chisel:bettishort}) can then be written as
\begin{align}
 &\int\limits_{-\infty}^\infty
			      \left\{\bR\av{\bU}(x_1-x_1')\av{\bs}^{(+)}(x_1)-\bR\av{\bS}(x_1-x_1')\av{\bu}(x_1)\right\}
			      \mathrm{d}x_1\nonumber
\\&\qquad=-\int\limits_{-\infty}^\infty
			      \left\{\bR\av{\bU}(x_1-x_1')\av{\bp}^{(-)}(x_1)+\frac{1}{4}\bR\jump{\bU}(x_1-x_1')\jump{\bp}^{(-)}(x_1)\right\}
			      \mathrm{d}x_1,
\end{align}
or alternatively in convolution notation (with convolution with respect to $x_1$): 
\begin{equation}\label{chisel:summativebetti}
 (\bR\av{\bU})^T*\av{\bs}^{(+)}-(\bR\av{\bS})^T*\av{\bu}=-(\bR\av{\bU})^T*\av{\bp}^{(-)}-\frac{1}{4}(\bR\jump{\bU})^T*\jump{\bp}^{(-)}.
\end{equation}
Applying the Fourier transform with respect to $x_1$, defined as
\begin{equation}
 \mathcal{F}\{f(x_1);\xi\}\equiv\bar{f}(\xi):=\int\limits_{-\infty}^\infty{f(x_1)e^{i\xi x_1}}\mathrm{d}x_1,
\end{equation}
to equation (\ref{chisel:summativebetti}), we arrive at the novel integral identity
\begin{equation}\label{chisel:newbetti}
 \av{\bar \bU}^T \bR \av{\bar \bs}^+  -  \av{\bar \bS}^T \bR \av{\bar \bu}   =   -\av{\bar\bU}^T\bR\av{\bar{\bp}}^-   -\frac{1}{4}\jump{\bar\bU}^T\bR\jump{\bar\bp}^-.
\end{equation}
Here and in what follows, the superscript $\pm$ which follow some Fourier transformed physical quantities denote the half-plane (upper or lower) in which the Fourier transform is analytic since the corresponding non-transformed entity is non-zero only on a half-line. Note however that in general, $\av{u}$ may be non-zero along both $x_1<0$ and $x_1>0$. This newly derived integral identity \eqref{chisel:newbetti} which results from summing Betti identities relates the mean displacement $\av{\bu}$ to the interfacial traction $\av{\bs}^{(+)}$ and crack face loadings $\av{\bp}$ and $\jump{\bp}$ and complements the identity found by Piccolroaz et al \cite{picc2009}:
\begin{equation}
 \jump{\bar \bU}^T \bR \av{\bar \bs}^+  -  \av{\bar \bS}^T \bR \jump{\bar \bu}   =   -\jump{\bar\bU}^T\bR\av{\bar{\bp}}   -\av{\bar\bU}^T\bR\jump{\bar\bp}.
\end{equation}

\section{Mode III}
In order to demonstrate the utility of identity \eqref{chisel:newbetti}, we first analyse out-of-plane deformation. Throughout, we will assume that
\begin{equation}
 \av{\sigma_{23}}^{(+)}(x_1)=O(x_1^{-1/2}),\quad x_1\to0;\qquad \av{\sigma_{23}}^{(+)}(x_1)=O(x_1^{-3/2}),\quad x_1\to+\infty,
\end{equation}
and so the Fourier transformed interfacial traction has asymptotic behaviour of the form
\begin{equation}
 \av{\bar\sigma_{23}}^+(\xi)=\sigma_0 + O(\xi^{1/2}),\quad \xi\to0;\qquad \av{\bar\sigma_{23}}^+(\xi)=O(\xi^{-1/2}),\quad \xi\to\pm\infty.
\end{equation}
Similarly, we make the following assumptions on the behaviour of $\av{p_3}^{(-)}$ and its Fourier transform:
\begin{eqnarray}
  \av{p_3}^{(-)}(x_1)=O(x_1^{-1/2}),\quad x_1\to0;\qquad \av{p_3}^{(-)}(x_1)=O(x_1^{-3/2}),\quad x_1\to-\infty,\\
  \av{\bar p_3}^-(\xi)=-\sigma_0 + O(\xi^{1/2}),\quad \xi\to0;\qquad \av{\bar p_3}^-(\xi)=O(\xi^{-1/2}),\quad \xi\to\pm\infty.
\end{eqnarray}
Here, the relationship between the leading constants in the behaviour of the Fourier transforms of $\av{\bar \sigma_{23}}^+$ and $\av{\bar p_3}^+$ follows immediately from balance conditions. Balance conditions also yield that $\jump{\bar p}^-(0)=0$, implying
\begin{equation}
 \jump{\bar p}^-(\xi)=O(\xi^\alpha),\quad \xi\to0^+,
\end{equation}
for some $\alpha>0$. We also assume that $\jump{\bar p}(\xi)\to0$ as $\xi\to\pm\infty$.

By a process entirely analogous to that described in the previous section, the Betti formula \eqref{chisel:summativebetti} relating displacement and traction  $u_3$ and $\sigma_{23}$ to the weight function $U_3$ and $\Sigma_{23}$ reduces to the scalar equation 
\begin{equation}
\av{U_3}*\av{\sigma_{23}}^{(+)}-\av{\Sigma_{23}}*\av{u_3}=-\av{U_3}*\av{p_3}^{(-)}-\frac{1}{4}\jump{U_3}*\jump{p_3}^{(-)},
\end{equation}
which upon applying Fourier transforms yields
\begin{equation}\label{chisel:FTd}
  \av{\bar{U}_3}(\xi)\av{\bar{\sigma}_{23}}^+(\xi) -\av{\bar\Sigma_{23}}(\xi)\av{\bar u_3}(\xi)=-\av{\bar U_3}(\xi)\av{\bar p_3}^-(\xi)-\frac{1}{4}\jump{\bar U_3}(\xi)\jump{\bar p_3}^-(\xi).
\end{equation}
Defining the constant $\eta=\frac{\mu_--\mu_+}{\mu_-+\mu_+}$ and multiplying (\ref{chisel:FTd}) throughout by $\eta\av{\bar U_3}^{-1}(\xi)$ gives
\begin{equation}\label{chisel:AandB}
\eta\av{\bar{\sigma}_{23}}^+(\xi) -B(\xi)\av{\bar u_3}(\xi)=-\eta\av{\bar p_3}^-(\xi)-A(\xi)\jump{\bar p_3}^-(\xi),
\end{equation}
where
\begin{equation}
 A(\xi)=\frac{\eta\jump{\bar U_3}(\xi)}{4\av{\bar U_3}(\xi)},\qquad B(\xi)=\frac{\eta\av{\bar\Sigma_{23}}(\xi)}{\av{\bar U_3}(\xi)}.
\end{equation}
Piccolroaz \cite{picc2009} gives the necessary relationships between these existing weight functions, namely
\begin{equation}
 \av{\bar U_3}(\xi)=\frac{\eta}{2}\jump{\bar U_3}(\xi),\qquad \jump{\bar U_3}(\xi)=-\frac{k}{|\xi|}\av{\bar\Sigma_{23}}(\xi),
\end{equation}
where $k$ is the bi-material constant
\begin{equation}
 k=\frac{\mu_++\mu_-}{\mu_+\mu_-}.
\end{equation}
It follows that $A(\xi)$ and $B(\xi)$ are given by
\begin{equation}
 A(\xi)=\frac{1}{2},\qquad B(\xi)= - \frac{2|\xi|}{k},
\end{equation}
whence (\ref{chisel:AandB}) becomes
\begin{equation}\label{chisel:id}
\av{\bar{\sigma}_{23}}^+(\xi)+\av{\bar p_3}^-(\xi) =-\frac{1}{2\eta}\jump{\bar p_3}^-(\xi)-\frac{2|\xi|}{k\eta}\av{\bar u_3}(\xi),
\end{equation}
which accompanies Piccolroaz's previous identity
\begin{equation}\label{chisel:idold}
 \av{\bar{\sigma}_{23}}^+(\xi)+\av{\bar p_3}^-(\xi) =-\frac{\eta}{2}\jump{\bar p_3}^-(\xi)-\frac{|\xi|}{k}\jump{\bar u_3}^-(\xi).
\end{equation}
Together, \eqref{chisel:id} and \eqref{chisel:idold} allow for the derivation of useful new integral identities which can be used to solve a much larger class of problems. Fundamentally they represent two equations in four key quantities, namely the Fourier transforms of: the sum of symmetric tractions along the crack and interface line $(\av{\bar{\sigma}_{23}}^+(\xi)+\av{\bar p_3}^-(\xi))$, the crack face skew-symmetric tractions, $\jump{\bar{p}_3}^-$, the crack face symmetric tractions, $\av{\bar{p}_3}^-$, and (novelly) the mean displacement, $\av{\bar{u}_3}$. Each equation relates three of these quantities; accordingly two more can be immediately derived, each of which neglect one quantity:
\begin{equation}\label{chisel:id3}
 \jump{\bar{p}_3}^-(\xi)=-\frac{1}{2}(\mu_++\mu_-)|\xi|\left(2\av{\bar{u}_3}(\xi)-\eta\jump{\bar{u}_3}^-(\xi)\right);
\end{equation}
\begin{equation}\label{chisel:id4}
\av{\bar{\sigma}_{23}}^+(\xi)+\av{\bar p_3}^-(\xi) =\frac{1}{2}(\mu_--\mu_+)|\xi|\left(\av{\bar{u}_3}(\xi)-\frac{1}{2\eta}\jump{\bar{u}_3}(\xi)\right).
\end{equation}
It is instructive to note that in general for a function $f^{(-)}$ with Fourier transform $\bar f^-$,
\begin{equation}
 \mathcal{F}^{-1}[|\xi|\bar{f}^-]=\frac{1}{\pi x_1}*\frac{\partial f^{(-)}}{\partial x_1}=\mathcal{S}\left\{\frac{\partial f^{(-)}}{\partial x_1}\right\}=\frac{1}{\pi}\int\limits_{-\infty}^\infty\frac{1}{x_1-t}\frac{\partial f^{(-)}}{\partial t}\mathrm{d}t.
\end{equation}
Here we have introduced the classic singular operator of the Cauchy type $\mathcal{S}$ to simplify notation in what follows, which transforms any H\"{o}lder continuous function into another H\"{o}lder continuous function \cite{muskhelishvili}. We further define the orthogonal projection operators $\mathcal{P}_\pm$ acting on $\mathbb{R}$:
\begin{equation}
 \mathcal{P}_\pm f = \begin{cases}
                      f(x_1)&\pm x_1\geq 0,\\
                      0&\text{otherwise},
                     \end{cases}
\end{equation}
and will define the singular operator $\mathcal{S}^{(s)}=\mathcal{P}_-\mathcal{S}\mathcal{P}_-$ and the compact operator $\mathcal{S}^{(c)}=\mathcal{P}_+\mathcal{S}\mathcal{P}_-$; further discussion of these operators, including the inversion of $\mathcal{S}^{(s)}$, can be found in \cite{picc2d3d}. With these operators now defined, taking inverse Fourier transforms of \eqref{chisel:idold} respectively with $x_1<0$ and $x_1>0$ yields Piccolroaz's identities
\begin{equation}\label{chisel:picc2009id1}
 \av{p_3}^{(-)}+\frac{\eta}{2}\jump{p_3}^{(-)}=-\frac{1}{k}\mathcal{S}^{(s)}\frac{\partial \jump{u_3}^{(-)}}{\partial x_1},\quad x_1<0,
\end{equation}
\begin{equation}\label{chisel:picc2009id2}
 \av{\sigma_{23}}^{(+)}=-\frac{1}{k}\mathcal{S}^{(c)}\frac{\partial \jump{u_3}^{(-)}}{\partial x_1},\quad x_1>0.
\end{equation}
Taking inverse Fourier transforms of \eqref{chisel:id} respectively with $x_1<0$ and $x_1>0$ yields the following novel identities, which include terms involving the mean displacement $\av{u_3}$:
\begin{equation}
 \av{p_3}^{(-)}+\frac{1}{2\eta}\jump{p_3}^{(-)}=-\frac{2}{k\eta}\mathcal{S}\frac{\partial \av{u_3}}{\partial x_1},\quad x_1<0,
\end{equation}
\begin{equation}
 \av{\sigma_{23}}^{(+)}=-\frac{2}{k\eta}\mathcal{S}\frac{\partial \av{u_3}}{\partial x_1},\quad x_1>0.
\end{equation}
Alternatively, \eqref{chisel:id} may be multiplied by $|\xi|^{-1}$ before Fourier inversion to give
\begin{equation}
 \mathcal{F}^{-1}_{x_1<0}\left\{\frac{\av{\bar{\sigma}_{23}}^++\av{\bar p_3}^-}{|\xi|}\right\}+\frac{1}{2\eta}\mathcal{F}_{x_1<0}^{-1}\left\{\frac{\jump{\bar p_3}^-}{|\xi|}\right\}=-\frac{2}{k\eta}\av{u_3}^{(-)},\quad x_1<0,
\end{equation}
\begin{equation}
 \mathcal{F}^{-1}_{x_1>0}\left\{\frac{\av{\bar{\sigma}_{23}}^++\av{\bar p_3}^-}{|\xi|}\right\}+\frac{1}{2\eta}\mathcal{F}_{x_1>0}^{-1}\left\{\frac{\jump{\bar p_3}^-}{|\xi|}\right\}=-\frac{2}{k\eta}\av{u_3}^{(+)},\quad x_1>0.
\end{equation}


Similarly, \eqref{chisel:id3} and \eqref{chisel:id4} yield the following four identities:
\begin{equation}
 \jump{p_3}^{(-)}=-(\mu_++\mu_-)\left(\mathcal{S}\frac{\partial\av{u_3}}{\partial x_1}-\frac{\eta}{2}\mathcal{S}^{(s)}\frac{\partial\jump{u_3}^{(-)}}{\partial x_1}\right),\quad x_1<0,
\end{equation}
\begin{equation}
0=-(\mu_++\mu_-)\left(\mathcal{S}\frac{\partial\av{u_3}}{\partial x_1}-\frac{\eta}{2}\mathcal{S}^{(c)}\frac{\partial\jump{u_3}^{(-)}}{\partial x_1}\right),\quad x_1>0.
\end{equation}
\begin{equation}
 \av{p_3}^{(-)}=\frac{1}{2}(\mu_--\mu_+)\left(\mathcal{S}\frac{\partial\av{u_3}}{\partial x_1}-\frac{1}{2\eta}\mathcal{S}^{(s)}\frac{\partial \jump{u_3}^{(-)}}{\partial x_1}\right),\quad x_1<0,
\end{equation}
\begin{equation}
 \av{\sigma_{23}}^{(+)}=\frac{1}{2}(\mu_--\mu_+)\left(\mathcal{S}\frac{\partial\av{u_3}}{\partial x_1}-\frac{1}{2\eta}\mathcal{S}^{(c)}\frac{\partial \jump{u_3}^{(-)}}{\partial x_1}\right),\quad x_1>0.
\end{equation}

\section{Applications}
The identity (\ref{chisel:id}) derived in the previous section can be coupled with the existing work of Piccolroaz et al \cite{picc2009} in order to solve two main classes of problems. One problem is the full determination of displacements of semi-infinite crack faces and bi-material interface for a crack subjected to given out-of-plane loadings. The existing identities in the literature allow only for the determination of the displacement jump across the crack faces; while this is a useful quantity it does not provide the full picture, since in general the crack opening will not be symmetric and so the average displacement is a quantity of interest which does not appear in the existing integral identities. Another problem is the determination of tractions on the crack faces from a given crack displacement profile. 

In this section, we will outline how the newly derived identity can solve both of these classes of problem and will illustrate with some numerical examples for the Mode III case.

\subsection{Crack profile determination for known loadings}
\subsubsection{Anti-symmetric loading}\label{subsection:antisym}
By means of example, let us consider the out-of-plane anti-symmetric loadings on the upper and lower crack faces given by
\begin{equation}
 p^{(-)}_\pm(x_1)=\pm \frac{F}{2}\left(1+\frac{2x_1}{3}\right)e^{x_1}\sqrt{-x_1} ,\quad x_1<0,
\end{equation}
implying that $\av{p_3}^{(-)}(x_1)\equiv0$ and $\jump{p_3}^{(-)}(x_1)=Fe^{x_1}\left(1+\frac{2x_1}{3}\right)\sqrt{-x_1}$. We seek the displacement jump $\jump{u_3}^{(-)}$ across the crack and the mean displacement $\av{u_3}$ along the crack line and also the interface ahead of the crack.

Both $\jump{u_3}^{(-)}(x_1)$ for $x_1<0$ and $\av{\sigma_{23}}^{(+)}(x_1)$ for $x_1>0$ can be immediately computed from Piccolroaz's identity:
\begin{align}
 \jump{u_3}^{(-)}(x_1)&=\frac{F \eta  k \sqrt{-x_1} \left(1+x_1 e^{x_1}  \text{Ei}(-x_1)\right)}{3 \pi },\quad &x_1<0,\\ \av{\sigma_{23}}^{(+)}(x_1)&=\frac{F \eta  \left(1 +2 x_1+x_1(2 x_1+3)  \text{Ei}(-x_1)e^{x_1}\right)}{6 \pi  \sqrt{x}},\quad &x_1>0,
\end{align}
where the function $\text{Ei}$ is the exponential integral function defined as 
\[
 \text{Ei}(x)=-\int\limits_{-x}^\infty \frac{e^{-t}}{t}\mathrm{d}t.
\]

We now detail how the identity \eqref{chisel:id3} can then be used to receive $\av{u_3}(x_1)$ for $x_1<0$. Rearranging \eqref{chisel:id3} gives
\begin{equation}\label{chisel:rearr3}
 \av{\bar{u}_3}(\xi)=\frac{\eta}{2}\jump{\bar u_3}^-(\xi)-\frac{1}{\mu_++\mu_-}\frac{\jump{\bar{p_3}}^-(\xi)}{|\xi|},
\end{equation}
which upon taking inverse Fourier transforms for $x_1<0$ yields
\begin{equation}\label{chisel:rearr3a}
 \av{u_3}^{(-)}(x_1)=\frac{\eta}{2}\jump{u_3}^{(-)}(x_1)-\frac{1}{\mu_++\mu_-}\mathcal{F}^{-1}_{x_1<0}\left\{ \frac{\jump{\bar{p_3}}^-(\xi)}{|\xi|} \right\},\quad x_1<0.
\end{equation}
For the given loading, the inverse Fourier transformed quantity evaluates to
\begin{equation}
\mathcal{F}^{-1}_{x_1<0}\left\{ \frac{\jump{\bar{p_3}}^-(\xi)}{|\xi|} \right\}=
-\frac{F \left(2\sqrt{\pi } e^{x_1}  x_1^{3/2} \text{erf}\left(\sqrt{x_1}\right)+2 x_1-1\right)}{3 \sqrt{\pi }},\quad x_1<0,
\end{equation}
where $\text{erf}(x)=\frac{2}{\sqrt{\pi}}\int\limits_0^z e^{-t^2}\mathrm{d}t$ denotes the error function, and so $\av{u}^{(-)}$ is now known. In order to calculate the mean out-of-plane displacement of the interface line (that is $\av{u}^{(+)}(x_1)$ for $x_1>0$), we can simply take inverse Fourier transforms of \eqref{chisel:rearr3} for $x_1>0$. Noting that $\jump{\bar{u}_3}^-$ is a minus-function analytic in the lower complex half-plane and so has inverse zero for $x_1>0$, this yields
\begin{equation}
 \av{u_3}^{(+)}(x_1)=-\frac{F \left(2 \sqrt{\pi } e^{x_1} {x_1}^{3/2} (1-\text{erf}\left(\sqrt{x_1}\right))-2 x_1+1\right)}{3 \sqrt{\pi }(\mu_++\mu_-)},\quad x_1>0.
\end{equation}
Thus $\av{u_3}$ and $\jump{u_3}^{(-)}$ are both determined and full information about the out-of-plane displacement profile of the crack has been obtained.

\subsubsection{Symmetric loading}
Let us now use the integral identities to determine the displacement profile of a symmetrically loaded crack. For the sake of example, consider the
\begin{equation}
 p^{(-)}_\pm(x_1)= Fx_1e^{x_1} ,\quad x_1<0,
\end{equation}
Piccolroaz's identity yields
\begin{align}
 \jump{u_3}^{(-)}(x_1)&=-\frac{F k}{\sqrt{\pi }} \left(2 (x_1-1) \mathfrak{F}\left(\sqrt{-x_1}\right)+\sqrt{-x_1}\right),&\quad x_1<0,\\
 \av{\sigma_{23}}^{(+)}(x_1)&= F e^{x_1} x_1 (1-\text{erf}\left(\sqrt{x_1}\right))+\frac{F(1-2x_1)}{2 \sqrt{\pi x_1}},&\quad x_1>0,
\end{align}
where $\mathfrak{F}$ denotes Dawson's function
\[
 \mathfrak{F}(x)=e^{-x^2}\int\limits_0^x e^{y^2}\mathrm{d}y.
\]

Identity \eqref{chisel:id4} then allows for $\av{u}^{(-)}$ to be obtained, since
\begin{align}
 \av{u_3}^{(-)}(x_1)&=\frac{2}{\mu_--\mu_+}\mathcal{F}_{x<0}^{-1}\left\{\frac{\av{\bar{\sigma}_{23}}+\av{\bar{p}_3}}{|\xi|}\right\}+\frac{1}{2\eta}\jump{u_3}^{(-)}\\
 &=\frac{i F (\mu_--\mu_+)}{2
   \sqrt{\pi } \mu_-\mu_+} \left(\sqrt{\pi } e^{x_1} (x_1-1) \text{erf}\left(\sqrt{x_1}\right)+\sqrt{x_1}\right), \quad x_1<0,
\end{align}
and similarly $\av{u_3}^{(+)}$ is given by
\begin{align}
 \av{u_3}^{(+)}(x_1)&=\frac{2}{\mu_--\mu_+}\mathcal{F}_{x>0}^{-1}\left\{\frac{\av{\bar{\sigma}_{23}}+\av{\bar{p}_3}}{|\xi|}\right\}=0
\end{align}
in the symmetric loading case.

\subsection{Recovering tractions from known displacements}
Returning to the example considered in subsection \ref{subsection:antisym}, suppose conversely that the crack face displacements $\av{u_3}^{(-)}$ and $\jump{u_3}^{(-)}$ are known and $\av{p_3}^{(-)}$ and $\jump{p_3}^{(-)}$ are sought. Rearranging \eqref{chisel:rearr3a} in this case gives a known expression for $\mathcal{F}^{-1}_{x_1<0}\{\jump{\bar{p}_3}^-(\xi)/|\xi|\}=g(x)$, say. The problem is then to recover $\jump{p_3}^{(-)}(x_1)$ from the known function $g(x)$.

In order to do this, let us define the auxiliary function $\psi(x_1)=\int\limits_{-\infty}^{x_1}\jump{p_3}^{(-)}(t)\mathrm{d}t$ for $x_1<0$ and note that for a general function $\varphi$ which is sufficiently well-behaved for all Fourier and inverse transforms in the below expression to exist, 
\begin{equation}
\mF{}{-1}{\frac{\mathcal{F}\left\{\varphi'\right\}}{|\xi|}}
=\mF{}{-1}{\sgn\xi}*\mF{}{-1}{\frac{\mathcal{F}\left\{\varphi'\right\}}{\xi}}
=-\frac{1}{\pi x}*\varphi(x).
\end{equation}
It follows that $\mathcal{S}^{(s)}\psi(x_1)=-g(x_1)$ for $x_1<0$ and so
\begin{equation}
 \jump{p_3}^{(-)}(x_1)
 =-\frac{\mathrm{d}}{\mathrm{d}x_1}\left(\psi(x_1)\right)
 =-\frac{\mathrm{d}}{\mathrm{d}x_1}\left(\left(\mathcal{S}^{(s)}\right)^{-1}g(x_1)\right).
\end{equation}
We refer the reader to Piccolroaz \cite{picc2d3d} for methods of inversion of the singular operator $\mathcal{S}^{(s)}$.

A similar approach can allow us to recover the symmetric loadings from known displacements $\av{u_3}$ and $\jump{u_3}$. Returning to the example considered in the previous subsection, by applying similar reasoning to that used above to identity \eqref{chisel:id4}, we obtain
\begin{equation}
 \av{p_3}^{(-)}(x_1)=-\frac{\mathrm{d}}{\mathrm{d}x_1}\left(\left(\mathcal{S}^{(s)}\right)^{-1}\left\{\frac{1}{2}(\mu_--\mu_+)\left(\av{\bar u_3}-\frac{1}{2\eta}\jump{\bar u_3}\right)\right\}\right).
\end{equation}

\section{Modes I and II}
For plane strain deformation, the additive Betti identity relating the physical displacement and stress vectors $\bu=[u_1,u_2]^T$ and $\bs=[\sigma_{21},\sigma_{22}]^T$ with the weight functions $\bU$, $\bS$, was derived in section \ref{section:betti} and is given by
\begin{equation}\label{chisel:boldbetti}
 \bR \av{\bU} * \av{\bs}^{(+)} - \bR \av{\bS} * \av{\bu} = -\bR\av{\bU} * \av{\bp}^{(-)}-\frac{1}{4}\bR\jump{\bU} * \jump{\bp}^{(-)}.
\end{equation}
%
Multiplying this identity throughout by $\bR^{-1}\av{\bar\bU}^{-T}$ gives
\begin{equation}\label{chisel:CDeqn}
 \av{\bar \bs}^+-\bC\av{\bar\bu}=-\av{\bar\bp} - \bD\jump{\bar\bp},
\end{equation}
where $\bC$ and $\bD$ are given by
\begin{equation}
 \bC=\bR^{-1}\av{\bar \bU}^{-T} \av{\bar \bS}^T\bR,\qquad \bD=\frac{1}{4}\bR^{-1}\av{\bar \bU}^{-T} \jump{\bar\bU}^T\bR.
\end{equation}
These matrices can be explicitly computed using relationships between the weight functions obtained by Antipov et al \cite{antipov1999} and Piccolroaz et al. \cite{picc2009}, which state that
\begin{equation}
 \jump{\bar \bU}=-\frac{1}{|\xi|}\left[b\bI-id\sgn(\xi)\bE\right]\av{\bar\bS},\quad \av{\bar\bU}=-\frac{b}{2|\xi|}\left[\alpha\bI-i\gamma\sgn(\xi)\bE\right]\av{\bar\bS},
\end{equation}
where $b$ is defined below, $d/b$ and $\alpha$ are the so-called Dundurs parameters, and $\gamma$ is a further bi-material constant:
\begin{equation}
 d=\frac{1-2\nu_+}{2\mu_+}-\frac{1-2\nu_-}{2\mu_-},\quad \alpha=\frac{\mu_-(1-\nu_+)-\mu_+(1-\nu_-)}{\mu_-(1-\nu_+)+\mu_+(1-\nu_-)},
\end{equation}
\begin{equation}
 b=\frac{1-\nu_+}{\mu_+}-\frac{1-\nu_-}{\mu_-},\qquad\gamma=\frac{\mu_-(1-2\nu_+)+\mu_+(1-2\nu_-)}{2\mu_-(1-\nu_+)+2\mu_+(1-\nu_-)}.
\end{equation}
It follows that
\begin{equation}
 \bC=-\frac{2|\xi|}{b(\alpha^2-\gamma^2)}\left\{ \alpha\bI+i\gamma\sgn(\xi)\bE \right\},
\end{equation}
\begin{equation}
 \bD=\frac{1}{2b(\alpha^2-\gamma^2)}\left\{(b\alpha-d\gamma)\bI+i(b\gamma-d\alpha)\sgn(\xi)\bE\right\}.
\end{equation}

\subsection{Crack profile determination for known loadings}
In this section, let us consider the symmetric and skew-symmetric parts of the crack loadings $\av{\bp}(\bx)$, $\jump{\bp}(\bx)$ to be known. Piccolroaz and Mishuris \cite{picc2d3d} derive integral equations relating the displacement jump to $\av{\bp}$ and $\jump{\bp}$, namely
\begin{align}
 \av{\bp}+\bmA^{(s)}\jump{\bp}&=\bmB^{(s)}\frac{\partial \jump{\bu}^{(-)}}{\partial x_1},\quad x_1<0,\label{chisel:matid1}\\
 \av{\bs}^{(+)}+\bmA^{(c)}\jump{\bp}&=\bmB^{(c)}\frac{\partial \jump{\bu}^{(-)}}{\partial x_1}, \quad x_1>0,\label{chisel:matid2}
\end{align}
where 
\begin{align}
 \bmA^{(s)}&=\frac{b}{2(b^2-d^2)}\left[(b\alpha-d\gamma)\bI +(d\alpha-b\gamma)\bE\mS^{(s)}\right],\\
 \bmB^{(s)}&=-\frac{1}{b^2-d^2}\left[b\bI\mS^{(s)} -d\bE\right],\\
 \bmA^{(c)}&=\frac{b(d\alpha-b\gamma)}{2(b^2-d^2)}\bE\mS^{(c)}, \qquad \bmB^{(c)}=-\frac{b}{b^2-d^2}\bI\mS^{(c)}.
\end{align}

Thus the singular matrix operator $\bmB^{(s)}$ may be inverted in equation (\ref{chisel:matid1}) to yield an expression for ${\partial\jump{\bu}}^{(-)}/{\partial x_1}$ and thus $\jump{\bu}(x_1,x_3)$ for $x_1<0$. Identity (\ref{chisel:matid2}) then gives an expression for $\av{\bs}^{(+)}$.

After splitting $\av{\bu}=\av{\bu}^{(+)}+\av{\bu}^{(-)}$ in the usual fashion and applying inverse Fourier transforms for $x_1<0$, the identity (\ref{chisel:CDeqn}) can be written as
\begin{equation}
 \av{\bu}^{(-)}=\mF{x<0}{-1}{\bC^{-1}\av{\bar\bs}+\bC^{-1}\av{\bar\bp}+\bC^{-1}\bD\jump{\bar\bp}},\quad x_1<0.
\end{equation}
Note that the inverse of the matrix $\bC$ is given by 
\begin{equation}
 \bC^{-1}(\xi)=-\frac{b}{2|\xi|}\{\alpha \bI -i\gamma\sgn(\xi)\bE\},
\end{equation}
while the matrix multiplying $\jump{\bp}$ is given by
\begin{equation}
 \bC^{-1}\bD=-\frac{1}{4|\xi|}\{b\bI-id\sgn(\xi)\bE\} ,
\end{equation}
and recall that balance conditions dictate that $\av{\bar\bs}(\mathbf{0})=-\av{\bar\bp}(\mathbf{0})$; in conjunction with the previously stated assumptions on the asymptotic behaviour of $\av{\bar\bs}$ and $\av{\bar\bp}$, this demonstrates that the inverse Fourier transform exists.

In order to find $\av{\bu}^{(+)}$, we can apply inverse Fourier transforms to (\ref{chisel:CDeqn}) for $x>0$, yielding
\begin{equation}
 \av{\bu}^{(+)}=\mF{x>0}{-1}{\bC^{-1}\av{\bar\bs}+\bC^{-1}\av{\bar\bp}+\bC^{-1}\bD\jump{\bar\bp}},\quad x_1>0.
\end{equation}

\subsection{Determination of crack face tractions for known displacement profile}
Let us now conversely suppose that the displacement profile of the crack is known and that we wish to determine the loadings on the crack faces.

The identity (\ref{chisel:matid1}) immediately yields the linear combination $\av{\bp}+\bmA^{(s)}\jump{\bp}$. Combining the Fourier transformed identities, we find that
\[
\jump{\bp}^{(-)}=\mathcal{F}_{x<0}^{-1}\left\{(\mathbf{A}-\mathbf{D})^{-1}(\mathbf{B}\jump{\bar\bu} - \mathbf{C}\av{\bar\bu})\right\}
\]
\[
 \av{\bp}^{(-)}=\mathcal{F}_{x<0}^{-1}\left\{(\mathbf{D}^{-1}-\mathbf{A}^{-1})^{-1}\left(\mathbf{D}^{-1}\mathbf{C}\av{\mathbf{\bar \bu}}-\mathbf{A}^{-1}\mathbf{B}\jump{\bar\bu}\right)\right\}
\]
or alternatively
\[
\jump{\bp}^{(-)}=\mathcal{F}_{x<0}^{-1}\left\{\mathbf{K}\jump{\bar\bu} - \mathbf{L}\av{\bar\bu}\right\}
\]
\[
 \av{\bp}^{(-)}=\mathcal{F}_{x<0}^{-1}\left\{\mathbf{K}\av{\mathbf{\bar \bu}}-\mathbf{M}\jump{\bar\bu}\right\}
\]
where the matrices $\mathbf{K},\mathbf{L},\mathbf{M}$ are given by
\begin{align}
 \mathbf{K}&=\beta_3\mathbf{I}|\xi|+i\beta_4\mathbf{E}\xi,\\
 \mathbf{L}&=4(\beta_1\mathbf{I}|\xi|-i\beta_2\mathbf{E}\xi);\\
 \mathbf{M}&=\beta_1\mathbf{I}|\xi|+i\beta_2\mathbf{E}\xi,
\end{align}
where
\begin{equation}
 \beta_1=\frac{\mu_+(\nu_+-1)}{4\nu_+-3}+\frac{\mu_-(\nu_--1)}{4\nu_--3};\qquad \beta_2=\frac{\mu_-(2\nu_--1)}{2(4\nu_--3)}-\frac{\mu_+(2\nu_+-1)}{2(4\nu_+-3)};
\end{equation}
\begin{equation}
 \beta_3=\frac{2\mu_-(\nu_--1)}{4\nu_--3}-\frac{2\mu_+(\nu_+-1)}{4\nu_+-3};\qquad 
 \beta_4=\frac{\mu_+(1-2\nu_+)}{4\nu_+-3}+\frac{\mu_-(1-2\nu_-)}{4\nu_--3}.
\end{equation}

\section{Conclusions}
While previously derived integral identities allowed for the displacement jump $\jump\bu$ across a crack to be determined for a given set of balanced but not necessarily symmetrical loadings, the derivation of integral identities which include the quantitity $\av{\bu}$ brings many benefits. Firstly, it allows the full displacement profile of the crack to be determined rather than solely the crack opening. Moreover, it allows for unknown loadings to be determined for known displacement profiles. The resulting identities have potential applications across a range of multiphysics problems, including but by no means limited to hydraulic fracture.

\section{Acknowledgements}
 A.V. gratefully acknowledges support from the European project FP7-PEOPLE-IAPP-284544-PARM-2.

\vspace{3em}

\noindent{\Large \bf Appendix}

\appendix
\section{Further example}
As a further illustrative example in which loadings are neither purely symmetric nor asymmetric, let us consider the upper and lower crack faces having respective loadings of
\begin{equation}\label{chisel:specificloadings}
 p^{(-)}_+(x)=xe^x, \quad p^{(-)}_-(x)=-e^x,\quad x<0,
\end{equation}
with material composition such that $\eta=2$ and $k=1$. By inverting the integral operator in identity (\ref{chisel:picc2009id1}) and then substituting the result into (\ref{chisel:picc2009id2}), we obtain the following expressions for the displacement jump and interfacial tractions:
\begin{align}
 \jump{u}(x)&=\frac{e^x \sqrt{-\pi x} (2-3x) \text{erfi}\left(\sqrt{-x}\right)+3 x}{2 \sqrt{-\pi x}},\quad x<0,\label{chisel:examplejump}\\
 \av{\sigma}(x)&=\frac{1}{4}\left(\frac{1-6x}{\sqrt{\pi x}}+2e^x(3x+1)(1-\text{erf}(\sqrt x))\right),\quad x>0.\label{chisel:examplesigma}
\end{align}
where $\text{erf}(x)=\frac{2}{\sqrt{\pi}}\int\limits_0^x e^{-t^2}\mathrm{d}t$ and 
$\text{erfi}(x)$ is the imaginary error function $\text{erfi}(x)=\text{erf}(iz)/i$ (which is real-valued for real arguments). Consequently, the Fourier transform of $\av{\sigma}$ is given by 
\begin{equation}\label{chisel:examplesigmabar}
\av{\bar\sigma}^+(\xi)=\frac{\left(2+\sqrt{-i \xi }\right)i \xi +7 \sqrt{-i \xi }-4}{4 (\xi -i)^2}.
\end{equation}
The inverse Fourier transform must be computed in order to recover the average displacement $\av{u}(x)$.
\begin{figure}[h]
 \includegraphics[width=\textwidth]{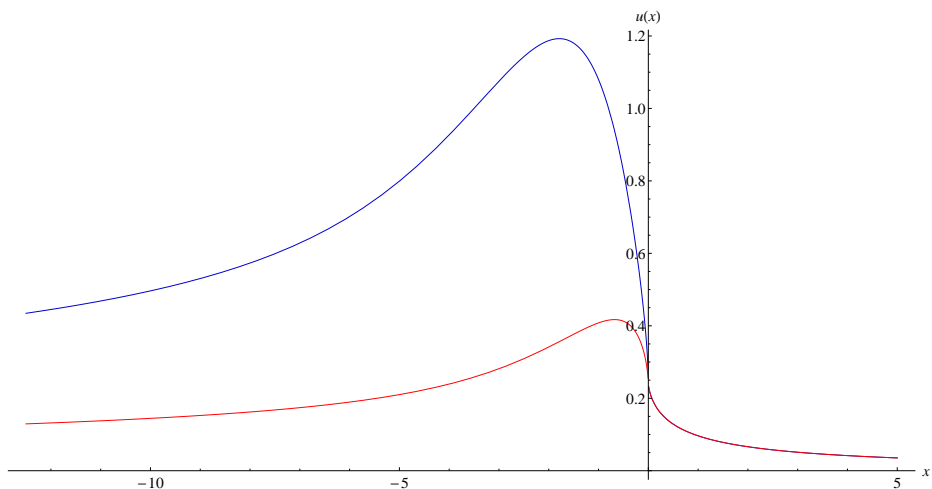}
 \caption{\footnotesize Out-of-plane crack face displacements for loading of the form given in (\ref{chisel:specificloadings})}\label{chisel:fig:dispprofile}
\end{figure}

\subsection{Determination of crack face tractions for known displacement profile}
Suppose, in contrast to the previously discussed formulation, that the crack displacement profile (that is, $\jump{u_3}(x)$ and $\av{u_3}(x)$ for $x<0$) is known but the tractions acting on the crack faces are sought. 

At first glance it may appear that multiplying identities \eqref{chisel:id3} and \eqref{chisel:id4} by $|\xi|^{-1}$ and applying the operator $\mathcal{F}^{-1}_{x<0}$ will immediately yield $\jump{p_3}$ and $\av{p_3}$. However, we assume that
\begin{equation}
 \av{\sigma_{23}}^{(+)}(x_1)=O(x_1^{-1/2}),\quad x_1\to0;\qquad \av{\sigma_{23}}^{(+)}(x_1)=O(x_1^{-3/2}),\quad x_1\to+\infty,
\end{equation}
and so the Fourier transformed interfacial traction has asymptotic behaviour of the form
\begin{equation}
 \av{\bar\sigma_{23}}^+(\xi)=\sigma_0 + O(\xi^{1/2}),\quad \xi\to0;\qquad \av{\bar\sigma_{23}}^+(\xi)=O(\xi^{-1/2}),\quad \xi\to\pm\infty.
\end{equation}
Similarly, we make the following assumptions on the behaviour of $\av{p_3}^{(-)}$ and its Fourier transform:
\begin{eqnarray}
  \av{p_3}^{(-)}(x_1)=O(x_1^{-1/2}),\quad x_1\to0;\qquad \av{p_3}^{(-)}(x_1)=O(x_1^{-3/2}),\quad x_1\to-\infty,\\
  \av{\bar p_3}^-(\xi)=-\sigma_0 + O(\xi^{1/2}),\quad \xi\to0;\qquad \av{\bar p_3}^-(\xi)=O(\xi^{-1/2}),\quad \xi\to\pm\infty.
\end{eqnarray}
Here, the relationship between the leading constants in the behaviour of the Fourier transforms of $\av{\bar \sigma_{23}}^+$ and $\av{\bar p_3}^+$ follows immediately from balance conditions. 
Consequently the inverse Fourier transforms of $\av{\bar p_3}/|\xi|$ and $\av{\bar \sigma_{23}}/|\xi|$ do not exist in the classical sense. 
Balance conditions also yield that $\jump{\bar p}^-(0)=0$, implying
\begin{equation}
 \jump{\bar p}^-(\xi)=O(\xi^\alpha),\quad \xi\to0^+,
\end{equation}
for some $\alpha>0$. We also assume that $\jump{\bar p}(\xi)\to0$ as $\xi\to\pm\infty$.

With these considerations in mind, we introducte the auxiliary functions
\begin{equation}
 \av{\bar p}_*^-(\xi)=\av{\bar p}^-(\xi)-\frac{\sigma_0}{(\xi-i)^2},\quad \av{\bar \sigma}_*^+(\xi)=\av{\bar \sigma}^-(\xi)+\frac{\sigma_0}{(\xi-i)^2},
\end{equation}
which have behaviour as $\xi\to0$ described by
\begin{equation}
 \av{\bar p}_*^-(\xi)=O(\xi^{1/2}),\quad \xi\to0;\qquad \av{\bar \sigma}_*^+(\xi)=O(\xi^{1/2}),\quad\xi\to0.
\end{equation}
In particular, we observe that 
\begin{eqnarray}
\mF{x<0}{-1}{\av{\bar p}_*^-(\xi)}=\av{p}^{(-)}(x)-\sigma_0 xe^x,\quad x<0,\\
\mF{x<0}{-1}{\av{\bar \sigma}_*^+(\xi)}=\av{\sigma}^{(+)}(x)+\sigma_0 xe^x,\quad x<0.
\end{eqnarray}


\begin{thebibliography}{}
\bibitem{antipov1999}{Antipov, Y.A.: An exact solution of the 3-D-problem of an interface semi-infinite plane crack. J. Mech.
Phys. Solids
47, 1051–1093 (1999)}

\bibitem{bueckner}{Bueckner, H. F., 1985. Weight functions and fundamental fields for the penny-shaped and the half plane crack in three-space.
Int. J. Solids Struct. 23, 57–93.}

\bibitem{mishurisv2014}{Mishuris, G., Piccolroaz, A., Vellender, A. (2014)
Boundary integral formulation for cracks at imperfect interfaces.
The Quarterly Journal of Mechanics and Applied Mathematics, 67, 363-387.}

\bibitem{morini2013}{Morini, L., Piccolroaz, A., Mishuris, G., Radi, E. (2013)
Integral identities for a semi-infinite interfacial crack in anisotropic elastic bimaterials.
International Journal of Solids and Structures, 50, 1437-1448.}

\bibitem{morini2015}{Morini, L., Piccolroaz, A. (2015) Boundary integral formulation for interfacial cracks in thermodiffusive bimaterials.
Proceedings of the Royal Society A, 471:20150284.}

\bibitem{muskhelishvili}{N.I. Mushkelishvili, Some Basic Problems of the Mathematical Theory of Elasticity, Noordhoff (1953)}

\bibitem{picc2009}{Piccolroaz, A., Mishuris, G., Movchan, A.B.: Symmetric and skew-symmetric weight functions in 2D
perturbation models for semi-infinite interfacial cracks. J. Mech. Phys. Solids
57, 1657–1682 (2009)}

\bibitem{picc2d3d}{Piccolroaz, A.,  Mishuris, G.: Integral Identities for a Semi-infinite Interfacial Crack in 2D and 3D Elasticity. J Elast {\bf 110}, 117--140 (2013).}




\bibitem{willismovchan}{Willis, J.R., Movchan, A.B.: Dynamic weight functions for a moving crack. I. Mode I loading. J. Mech. Phys. Solids 43, 319–341 (1995)}
\end{thebibliography}
\end{document}